\documentclass[iop]{emulateapj}
\usepackage[]{hyperref} 
\usepackage{color}
\usepackage[utf8]{inputenc}
\usepackage{graphicx,url,times}
\usepackage{multirow} 
\usepackage{booktabs,threeparttable}
\usepackage{amsmath}
\newif\ifpdf
  \ifx\pdfoutput\undefined
    \pdffalse
  \else
    \ifnum\pdfoutput=1
      \pdftrue
    \else
      \pdffalse
    \fi
  \fi

\ifpdf
    \graphicspath{{plots/}}
\else
    \graphicspath{{eps/}}
\fi

\newcommand{\papertitle}{Solving the puzzle of subhalo spins}
\ifpdf\hypersetup{
pdftitle={\papertitle},
pdfauthor={Yang Wang},
pdfkeywords={N-body simulations, haloes evolution, dark matter},
pdfstartview=FitH,
}
\fi

\newcommand{\ahf}{\textsc{ahf}}
\newcommand{\hbt}{\textsc{hbt}}
\newcommand{\subfind}{\textsc{subfind}}

\newcommand{\halos}{haloes}
\newcommand{\Halos}{Haloes}
\newcommand{\hkpc}{{\ifmmode{h^{-1}{\rm kpc}}\else{$h^{-1}$kpc}\fi}}
\newcommand{\abs}[1]{\left|#1\right|}
\newcommand{\ghalo}{\textsc{ghalo}}
\newcommand{\Fig}[1]{Figure~\ref{#1}}
\newcommand{\Eqn}[1]{Equation~\ref{#1}}
\newcommand{\Tbl}[1]{Table~\ref{#1}}

\newcommand{\Sec}[1]{Section~\ref{#1}}
\newcommand{\Ssec}[1]{Subsection~\ref{#1}}

\newlength{\figwidth}
\newlength{\resplot}
\slugcomment{Accepted on 14th Jan, 2015}
\shortauthors{Wang et al.}

\begin{document}
\title {\papertitle}
\author{
Yang Wang\altaffilmark{1,2,4},
Weipeng Lin\altaffilmark{1,3},
Frazer R. Pearce\altaffilmark{2},
Hanni Lux\altaffilmark{2,5},
Stuart I. Muldrew\altaffilmark{6} \&
Julian Onions\altaffilmark{2} }
\altaffiltext{1}  {Key Laboratory for Research in Galaxies and Cosmology, Shanghai Astronomical Observatory, Shanghai 200030, China}
\altaffiltext{2}  {School of Physics \& Astronomy, University of Nottingham, Nottingham, NG7 2RD, UK}
\altaffiltext{3}  {School of Astronomy and Space Science, Sun Yat-Sen University, Guangzhou, 510275, China}
\altaffiltext{4}  {Graduate School of the Chinese Academy of Science, 19A, Yuquan Road, Beijing, China}
\altaffiltext{5}  {Department of Physics, University of Oxford, Denys Wilkinson Building, Keble Road, Oxford, OX1 3RH, UK}
\altaffiltext{6} {Department of Physics and Astronomy, University of Leicester, University Road, Leicester, LE1 7RH, UK}
\email{wangyang,linwp@shao.ac.cn}

\begin{abstract}
Investigating the spin parameter distribution of sub\halos\ in two
high resolution isolated halo simulations, recent work by Onions et
al. suggested that typical subhalo spins are consistently lower than
the spin distribution found for field \halos. To further examine this
puzzle, we have analyzed simulations of a cosmological volume with
sufficient resolution to resolve a significant subhalo population. We
confirm the result of Onions et al. and show that the typical spin of
a subhalo decreases with decreasing mass and increasing proximity to
the host halo center. We interpret this as the growing influence of
tidal stripping in removing the outer layers, and hence the higher
angular momentum particles, of the sub\halos\ as they move within the
host potential. Investigating the redshift dependence of this effect,
we find that the typical subhalo spin is smaller with decreasing
redshift.  This indicates a temporal evolution as expected in the
tidal stripping scenario.
\end{abstract}

\keywords{
methods: numerical --
galaxies: \halos --
galaxies: evolution --
cosmology: theory -- dark matter
}

\section{Introduction} \label{sec:introduction}
In the standard model of structure formation, the rotation velocities
of disc galaxies are correlated with the spin properties of their
surrounding dark matter \halos\ \citep{fall_1980}. The simplest model
explains this correlation via angular momentum conservation and assuming
baryons and dark matter initially share the same specific angular
momentum distribution \citep{mestel_1963}. Even though subsequent
models paint a more complex picture, this link continues to exist
\citep[e.g.][]{dalcanton_1997, mo_1998, navarro_2000,
abadi_2003,bett_2010}. Consequently, the halo spin is an important
parameter in many semi-analytic models of galaxy formation
\citep{kauffmann_1993,kauffmann_1997,frenk_1997,cole_2000,benson_2001,
bower_2006, croton_2006, delucia_2007, bertone_2007, font_2008,
benson_2012} and a number of studies have investigated the spin of
individual dark matter \halos\ in cosmological simulations
\citep{peebles_1969,bullock_2001,hetznecker_2006,bett_2007,maccio_2007,
gottlober_2007,knebe_2008,antonuccio_2010,wang_2011b, lacerna_2012,
trowland_2012, bryan_2013}.

Due to a lack of resolution in previous generations of large
cosmological simulations, subhalo spins have not been thoroughly
investigated so far, despite their application within current
semi-analytic models \citep{guo_2011}. Initial work by
\citet{lee_2013} analyzed the spins of the two most massive
substructures of Local Group like systems in the Millennium-II
simulation \citep{boylan_kolchin_2009} and revealed possible
consequences for the application of subhalo spins to near-field
cosmology.

\citet{onions_2013} investigated the spin distribution of sub\halos\
in two high resolution simulations of a Milky Way-like halo (the
Aquarius simulation \citep{springel_aquarius_2008} and the \ghalo\
simulation \citep{stadel_ghalo_2009}) analyzed by a variety of subhalo
finders.  They suggested that subhalo spins are significantly offset
to lower values than those seen in typical distribution functions
fitted to \halos\ \citep{bullock_2001, bett_2007}.  As this result is
independent of the subhalo finder used, it suggests that this is a
true physical effect.  This could not be investigated further because
their resimulation did not contain a large field halo population.
Excluding sub\halos, \citet{colin_2004} found the spin parameter
distribution of isolated dwarf dark matter \halos\ to be perfectly
consistent with that of larger \halos.  This suggests that the
consistently lower spin of substructure is not due to the generally
smaller mass of sub\halos, but is more likely related to tidal
stripping of high angular momentum material. On the other hand, this
offset could also be due to differences between the Aquarius and
\ghalo\ simulations and those used by Bullock to define the field
relation. To answer this question we require a single
simulation that simultaneously includes both a significant subhalo and
field halo population.

In this work we use purpose built simulations, specifically designed to contain
both a field and subhalo population, to investigate the difference in
spin distribution functions between sub\halos\ and \halos.  In
\Sec{sec:data}, we present these simulations and the corresponding
(sub-)halo catalogues. The different theoretical models of
dimensionless spin parameters are described in \Sec{sec:theory}, while
our results are summarized in \Sec{sec:results}. We discuss our work
and conclude in \Sec{sec:discussion}.

\section{Simulation Data} \label{sec:data}
As we require our halo and subhalo masses to span a wide dynamic range
($10^{8}\lesssim\,M\,\lesssim10^{15}{\rm M_\odot}$) , we have run four
dark matter only comoving cosmological boxes containing $512^3$
particles, with linear sizes of $8\,h^{-1}{\rm Mpc}$, $20\,h^{-1}{\rm
Mpc}$, $50\,h^{-1}{\rm Mpc}$, $100\,h^{-1}{\rm Mpc}$ respectively
(hereafter BoxA, BoxB, BoxC, BoxD). The softening lengths are chosen
to be $4\%$ of the mean separation between particles. This set of
simulations can both sufficiently resolve subhalo spins \citep[at
least 300 particles per subhalo;][]{bett_2007} and have significant
statistics for \halos\ (c.f. \Tbl{tab:haloes}). We also ran two
simulations with the same parameters as those used for the BoxA
simulation except for the gravitational softening parameter. BoxA\_S1
has a smaller softening length whereas BoxA\_S2 has a larger softening
length. We also ran a low resolution simulation containing $256^3$
particles and the same linear size, $8\,h^{-1}{\rm Mpc}$ as BoxA which
we designate BoxLo. The mass resolution of BoxA ($2.6\times
10^5\,h^{-1}{\rm M_\odot}$ per particle) is very close to that of the
Aquarius-A simulation at level 4 \citep{springel_aquarius_2008} which
had a particle mass of $2.7\times 10^5\,h^{-1}{\rm M_\odot}$ in the
high resolution region.  This is roughly three times better mass
resolution than the Millennium-II simulation ($m_p = 6.9\times
10^6\,h^{-1}{\rm M_\odot}$) used by \citet{lee_2013}. The cosmology
was chosen to be the same as in the Aquarius simulation,
i.e. $\Lambda$CDM with $\Omega_M = 0.25,\: \Omega_\Lambda = 0.75,\:
\sigma_8 =0.9,\: n_s = 1,\: h = 0.73$.  Initial conditions were
generated at $z=127$ by the code N-GenIC using
 the Zel'dovich approximation (written by Volker Springel) to linearly
evolve positions from an initially glass-like state.  This was then
evolved to the present day using \textsc{gadget}-2
\citep{springel_cosmological_2005}.

All simulations except BoxLo were analyzed with the (sub-)halo finding code
\subfind\ \citep{subfind_2001}, \ahf\ \citep{gill_evolution_2004, knollmann_ahf:_2009} and \hbt\
\citep{han_resolving_2011}. BoxLo was only analyzed with \subfind\ .
A summary of our simulations is given in
\Tbl{tab:haloes} and details of all the halo finding algorithms we
have used and a discussion of their relative merits can be found in
\citet{knebe_stateofaffairs_2013}.

\begin{table*}
  \caption{Summary of simulation properties, halo and subhalo counts for the halo finder indicated.}
\label{tab:haloes}
\centering
\begin{threeparttable}
\begin{tabular}{ l  c c c c c c c c c}
\toprule
{Name} & {Box size} & {Particle mass} & {Force softening} & $N_{halo,\subfind}$ & $N_{halo,\hbt}$ & $N_{halo,\ahf}$ & {$N_{sub,\subfind}$} & {$N_{sub,\hbt}$} & {$N_{sub,\ahf}$} \\
            & {$h^{-1}{\rm Mpc}$}      & {$h^{-1}{\rm M_\odot}$} &{$h^{-1}{\rm kpc}$}& $\ge300$ & $\ge300$ & $\ge300$ & $\ge300$ & $\ge300$ & $\ge300$ \\
\midrule
BoxLo   &8  &$2.1\times10^{6}$ &1.25 &1136  &-    &-    &213  &-   &-   \\
\\
BoxA\_S1&8  &$2.6\times10^{5}$ &0.04 &6589  &6698 &6775 &1934 &2169&1460\\
BoxA    &8  &$2.6\times10^{5}$ &0.63 &6651  &6587 &6798 &1651 &1899&1216\\
BoxA\_S2&8  &$2.6\times10^{5}$ &1.25 &6585  &6476 &6529 &1388 &1618&944 \\
\\
BoxB    &20 &$4.1\times10^{6}$ &1.56 &8923  &8785 &9139 &2111 &2494&1302\\
BoxC    &50 &$6.5\times10^{7}$ &3.91 &12791 &12533&12874&2687 &3325&1597\\
BoxD    &100&$5.2\times10^{8}$ &7.81 &17562 &17053&16901&3072 &3949&1737\\
\bottomrule
&&&&$\ge2400$ & $\ge2400$ &$\ge2400$ & $\ge2400$ & $\ge2400$ & $\ge2400$ \\
\midrule
BoxA\tnote{*}&&&&                     1132  &-&-&215&-&-\\
\bottomrule
\end{tabular}
\begin{tablenotes}
\footnotesize
\item[*] This higher particle number threshold is used to compare with BoxLo over the same halo mass range.
\end{tablenotes}
\end{threeparttable}
\end{table*}

\section{Theory}\label{sec:theory}
The dimensionless spin parameter indicates how much a collection of
particles is supported by the angular momentum against gravitational collapse
assuming gravitational equilibrium, where a negligible spin
parameter represents minimal support, while the value for a
completely supported system depends on the chosen
parametrisation. There are two standard parametrisations defined
by \citet{peebles_1969} and \citet{bullock_2001}, respectively, that
we describe in the following two sections.

 \citet{hetznecker_2006} showed that Bullock's parametrisation is less
 dependent on redshift evolution than Peebles'
 parametrisation. This is due to it being more robust to variations in
 the position of the structure's outer radius and therefore not as
 strongly affected by the many minor mergers over a halo's merging
 history. Therefore, the two descriptions are not readily interchangeable
 and results need to be compared using the same parameter.

\subsection{Peebles Spin Parameter}\label{sec:peebles}
\cite{peebles_1969} proposed to parameterise the (sub-)halo spin in the following way:
\begin{equation}
\lambda = \frac{J \sqrt{\abs{E}}}{GM^{5/2}},
 \label{eqn:peebles}
\end{equation}
where $J$ is total angular momentum, $E$ the energy and $M$ the mass of the (sub-)structure. With this choice a value of  $\lambda \simeq 0.4$ represents a purely rotationally supported object \citep{frenk_2012}.

Applying this parametrisation \citet{bett_2007} determined the spin distribution of \halos\ in the
Millennium Simulation \citep{springel_millenium_2005}.  The Millennium Simulation has a mass
resolution of $m_p= 8.6\times10^8\,h^{-1}{\rm M_\odot}$ and therefore
contains very few sub\halos. The vast majority of the objects in the
TREEclean catalogue of \citet{bett_2007} are \halos\ rather than
sub\halos.  The probability density function of $\log\lambda$ they
found to fit the distribution used the following parametrisation;
\begin{equation}
P(\log \lambda) = A \left( \frac{\lambda}{\lambda_0} \right) ^3
\exp \left[ -\alpha \left(\frac{\lambda}{\lambda_0}\right)^{3/\alpha} \right]
 \label{eqn:bettfit}
\end{equation}
\noindent where A is given by,
\begin{equation}
 A = 3 \ln 10 \frac{\alpha^{\alpha-1}}{\Gamma(\alpha)},
\label{eqn:bettA}
\end{equation}
\noindent and $\Gamma(\alpha)$ is the gamma function.  They found $\lambda_0 = 0.04326$ and $\alpha=2.509$ best fit the distribution of halo spins.

\subsection{Bullock Spin Parameter}\label{sec:bullock}
\citet{bullock_2001} proposed a dimensionless spin parameter of the form:
\begin{equation}
  \lambda' = \frac{J}{\sqrt{2}MRV},
 \label{eqn:bullock}
\end{equation}
\noindent where $J$ is the angular momentum within a virilized sphere with
radius $R$ and mass $M$, and $V$ is the circular velocity at the
virial radius ($V^2=GM/R$). They also proposed a parametrisation of the probability density function based on \citet{barnes_1987},
\begin{equation}
 P(\lambda') = \frac{1}{\lambda' \sqrt{2\pi} \sigma}
\exp \left( - \frac{ \ln^2 (\lambda' / \lambda_0')}{ {2\sigma^2}}  \right).
\label{eqn:bullockfit}
\end{equation}
\noindent They found the best fit for \halos\ is given by
$\lambda_0' = 0.035$ and $\sigma = 0.5$.

\section{Results}\label{sec:results}
Note that in this section we will show results for the discrete,
normalized derivative of the spin distribution function $\Delta
N(<\log{\lambda})/\Delta\log{\lambda}/N_{tot}$ and $\Delta
N(<\lambda')/\Delta\lambda'/N_{total}$, while the fitted functions are
for the continuous probability density function
$P(\log\lambda)=dN(<\log{\lambda})/d\log{\lambda}/N_{total}$ and
$P(\lambda')=dN(<\lambda')/d\lambda'/N_{total}$, respectively.
For Peebles spin, we set the bin width to be $\Delta\log{\lambda}=(\log{\lambda_{max}}-\log{\lambda_{min}})/100$.
For Bullock spin, the bin width is $\Delta\lambda'=(\lambda'_{max}-\lambda'_{min})/100$.
\subsection{Halo finding code, softening, and resolution test}
\label{Ssec:Test}
In this section, we first test whether simulation data set and the
specific choice of substructure finding code will affect the derived
spin of (sub-)\halos. \citet{maccio_2008} has tested halo spins with
different cosmological parameters. They found that the spin
distributions of \halos\ is essentially independent of cosmology, at
least for changes between WMAP1,WMAP3 and WMAP5. We choose not to
confirm this result here. BoxA, BoxA\_S1 and BoxA\_S2 are used to
compare different force resolutions. BoxA and BoxLo are used to
compare different mass resolutions. To ensure reliable properties are
recovered only (sub-)\halos\ with more than 300 particles are selected
throughout this work. While calculating the spin of \halos, all their
substructures are removed from them. Spin distributions are fitted by
\Eqn{eqn:bettfit} and \Eqn{eqn:bullockfit}, and the fitting parameters
are listed in \Tbl{tab:codes}.

    \begin{table*}
     \caption{Parameters for the spin distribution with different substructure finding codes and force resolution.}
    \label{tab:codes}
    \centering
    \begin{tabular}{ l l c c c c}
    \toprule
    \multicolumn{2}{c}{\multirow{2}{*}{Peebles Spin}}&\multicolumn{2}{c}{\halos\ } &\multicolumn{2}{c}{sub\halos\ }\\
    && $\lambda_0$ & $\alpha$ & $\lambda_0$ & $\alpha$ \\
    \hline
    \multirow{3}{*}{\subfind}& BoxA\_S1 &$0.0398\pm0.00019 $&$ 2.54\pm0.039 $&$ 0.0237\pm0.00032 $&$ 3.56\pm0.13$\\
                             & BoxA     &$0.0371\pm0.00021 $&$ 2.59\pm0.047 $&$ 0.0254\pm0.00049 $&$ 2.92\pm0.17$\\
                             & BoxA\_S2 &$0.0364\pm0.00026 $&$ 2.59\pm0.058 $&$ 0.0298\pm0.00036 $&$ 2.60\pm0.10$\\
                             \hline
    \multirow{3}{*}{\hbt}    & BoxA\_S1 &$0.0390\pm0.00024 $&$ 2.42\pm0.049 $&$ 0.0260\pm0.00029 $&$ 2.89\pm0.10$\\
                             & BoxA     &$0.0366\pm0.00019 $&$ 2.48\pm0.042 $&$ 0.0277\pm0.00027 $&$ 2.68\pm0.08$\\
                             & BoxA\_S2 &$0.0356\pm0.00021 $&$ 2.37\pm0.046 $&$ 0.0322\pm0.00033 $&$ 2.52\pm0.08$\\
                             \hline
    \multirow{3}{*}{\ahf}    & BoxA\_S1 &$0.0380\pm0.00024 $&$ 2.82\pm0.053 $&$ 0.0303\pm0.00044 $&$ 2.53\pm0.12$\\
                             & BoxA     &$0.0369\pm0.00026 $&$ 2.76\pm0.059 $&$ 0.0343\pm0.00045 $&$ 2.84\pm0.11$\\
                             & BoxA\_S2 &$0.0367\pm0.00023 $&$ 2.61\pm0.051 $&$ 0.0384\pm0.00074 $&$ 3.09\pm0.17$\\
    \midrule
    \multicolumn{2}{c}{\multirow{2}{*}{Bullock Spin}}&\multicolumn{2}{c}{\halos\ } &\multicolumn{2}{c}{sub\halos\ }\\
    && $\lambda_0'$ & $\sigma$ & $\lambda_0'$ & $\sigma$ \\
    \hline
    \multirow{3}{*}{\subfind}& BoxA\_S1 &$0.0308\pm0.00027 $&$ 0.655\pm0.007 $&$ 0.0123\pm0.00017$&$ 0.827\pm0.009$\\
                             & BoxA     &$0.0310\pm0.00025 $&$ 0.629\pm0.007 $&$ 0.0167\pm0.00019 $&$ 0.664\pm0.009$\\
                             & BoxA\_S2 &$0.0320\pm0.00022 $&$ 0.615\pm0.006 $&$ 0.0224\pm0.00028 $&$ 0.629\pm0.010$\\
                             \hline
    \multirow{3}{*}{\hbt}    & BoxA\_S1 &$0.0303\pm0.00026 $&$ 0.638\pm0.007 $&$ 0.0150\pm0.00013 $&$ 0.754\pm0.007$\\
                             & BoxA     &$0.0309\pm0.00026 $&$ 0.626\pm0.007 $&$ 0.0198\pm0.00018 $&$ 0.674\pm0.008$\\
                             & BoxA\_S2 &$0.0320\pm0.00025 $&$ 0.605\pm0.007 $&$ 0.0262\pm0.00031 $&$ 0.644\pm0.010$\\
                             \hline
    \multirow{3}{*}{\ahf}    & BoxA\_S1 &$0.0278\pm0.00025 $&$ 0.677\pm0.008 $&$ 0.0178\pm0.00028 $&$ 0.768\pm0.013$\\
                             & BoxA     &$0.0283\pm0.00019 $&$ 0.655\pm0.006 $&$ 0.0248\pm0.00046 $&$ 0.761\pm0.016$\\
                             & BoxA\_S2 &$0.0290\pm0.00022 $&$ 0.613\pm0.006 $&$ 0.0323\pm0.00059 $&$ 0.779\pm0.015$\\
    \bottomrule
    \end{tabular}
    \end{table*}

As \Tbl{tab:codes} shows, the recovered spin properties of \halos\ are
largely independent of the choice of gravitational softening. For
sub\halos\ there is a slight trend for the Peebles spin parameter to
increase as the softening is increased but this effect is only barely
resolved. Such a trend would be expected as a larger softening will
produce a shallower core potential, lowering slightly the central
kinetic energy and altering the energetics and angular momentum
profile, thus affecting the spin parameter.  For the Peebles measure
(\Eqn{eqn:peebles}), the change in energy is outweighed by the change
in angular momentum but the two effects counteract each other.  For
the Bullock spin parameter (\Eqn{eqn:bullock}), only angular momentum
has an affect on the spins. So we find that the Bullock spin of
sub\halos\ is more sensitive to the softening, as shown in
\Tbl{tab:codes}.

\Tbl{tab:codes} also shows that, contrary to \cite{onions_2013}, the
three halo finding methods do not recover consistent spin
parameters. While they all agree on the halo spins, \ahf\ recovers
significantly larger spins on average for the subhalo population than
either \subfind\ or \hbt\ which are consistent with each other. The
subhalo spins for \ahf\ are broadly consistent with the field
population, particularly for larger values of the gravitational
softening. This is also discrepant with \cite{onions_2013} who found
lower spin parameter values for their sub\halos. This result is due to
the failure of \ahf\ to resolve a significant fraction of sub\halos\
within the simulations. The subhalo numbers given in \Tbl{tab:haloes}
indicate that around $36\%$ of the sub\halos\ containing 300 or more
particles in BoxA are missed by \ahf. Difficulties for \ahf\ in
resolving substructures where the density contrast between the subhalo
and the main halo is expected to be small have also been reported
elsewhere \citep{Avila_2014}. Further evidence for this issue is the
rising incidence of missing substructures as the box size is increased
evidenced in \Tbl{tab:haloes}: for the largest box (BoxD), \ahf\
misses $71\%$ of the sub\halos\ found by \subfind. \ahf\ is missing small
sub\halos\ in the outskirts of the host halo and (as we shall
demonstrate later) these small sub\halos\ are precisely the ones with
the lowest spin parameters.

This naturally
produces a population of sub\halos\ with higher average spin parameter
for \ahf. As \hbt\ and \subfind\ produce consistent results and \ahf\ fails
to recover the complete subhalo population we choose to concentrate our
analysis on \subfind\ for the remainder of this paper.

    \begin{table*}
     \caption{Parameters recovered by \subfind\ for the spin distribution with different mass resolution.}
    \centering
    \label{tab:res}
    \begin{tabular}{ l c c c c}
    \toprule
    \multirow{2}{*}{Peebles Spin}&\multicolumn{2}{c}{\halos\ } &\multicolumn{2}{c}{sub\halos\ }\\
    & $\lambda_0$ & $\alpha$ & $\lambda_0$ & $\alpha$ \\
    \hline
    BoxA($N_p\ge300$)     &$ 0.0371\pm0.00021 $&$ 2.59\pm0.047 $&$ 0.0254\pm0.00049 $&$ 2.92\pm0.17$\\
    BoxA($N_p\ge2400$)    &$ 0.0393\pm0.00062  $&$ 2.53\pm0.13  $&$ 0.0281\pm0.0011  $&$ 2.78\pm0.33$\\
    BoxA\_S2($N_p\ge2400$)&$ 0.0389\pm0.00069  $&$ 2.48\pm0.14  $&$ 0.0288\pm0.0010  $&$ 2.51\pm0.29$\\
    BoxLo($N_p\ge300$)    &$ 0.0392\pm0.00073  $&$ 2.38\pm0.14  $&$ 0.0293\pm0.0008  $&$ 2.12\pm0.23$\\
    \midrule
    \multirow{2}{*}{Bullock Spin}&\multicolumn{2}{c}{\halos\ } &\multicolumn{2}{c}{sub\halos\ }\\
    & $\lambda_0'$ & $\sigma$ & $\lambda_0'$ & $\sigma$ \\
    \hline
    BoxA($N_p\ge300$)      &$ 0.0310\pm0.00024 $&$ 0.629\pm0.007 $&$ 0.0167\pm0.00019 $&$ 0.664\pm0.009$\\
    BoxA($N_p\ge2400$)     &$ 0.0324\pm0.00060 $&$ 0.626\pm0.017 $&$ 0.0183\pm0.0008 $&$ 0.738\pm0.035$\\
    BoxA\_S2($N_p\ge2400$) &$ 0.0327\pm0.00056 $&$ 0.622\pm0.016 $&$ 0.0198\pm0.0009 $&$ 0.662\pm0.040$\\
    BoxLo($N_p\ge300$)     &$ 0.0337\pm0.00049 $&$ 0.616\pm0.012 $&$ 0.0207\pm0.0007 $&$ 0.582\pm0.028$\\
    \bottomrule
    \end{tabular}
    \end{table*}

In \Tbl{tab:res} we test the influence of mass resolution. In the
brackets after simulation name, we note the particle number threshold
chosen. This is set in order to match (sub)halo masses between BoxA
and BoxLo, the lower resolution version of this simulation. This
ensures that the halo and subhalo catalogues for BoxLo($N_p\ge300$),
BoxA($N_p\ge2400$) and BoxA\_S2($N_p\ge2400$) have the same mass
range. The gravitational softening lengths for both BoxLo and BoxA
were set to $4\%$ of mean particle separation. BoxLo and BoxA\_S2 have
the same absolute softening length ($1.25h^{-1}{\rm kpc}$). The results
show that mass resolution has almost no effect on spin
distribution. The first three rows of each sub-table also give a hint
about the influence of softening: as suggested above softening mainly
affects the spin of small sub\halos. This is not surprising for the
reasons already indicated.

\subsection{\Halos\ vs. Sub\halos}
\label{Ssec:HvsS}
So far we have seen that the lower spin parameter distribution
observed for sub\halos\ appears to be a robust result that does not
depend upon the choice of halo finder, gravitational softening or mass
resolution. Here we explore a possible physical origin for the lower
subhalo spins.

    \begin{figure}
    \includegraphics[width=1\linewidth]{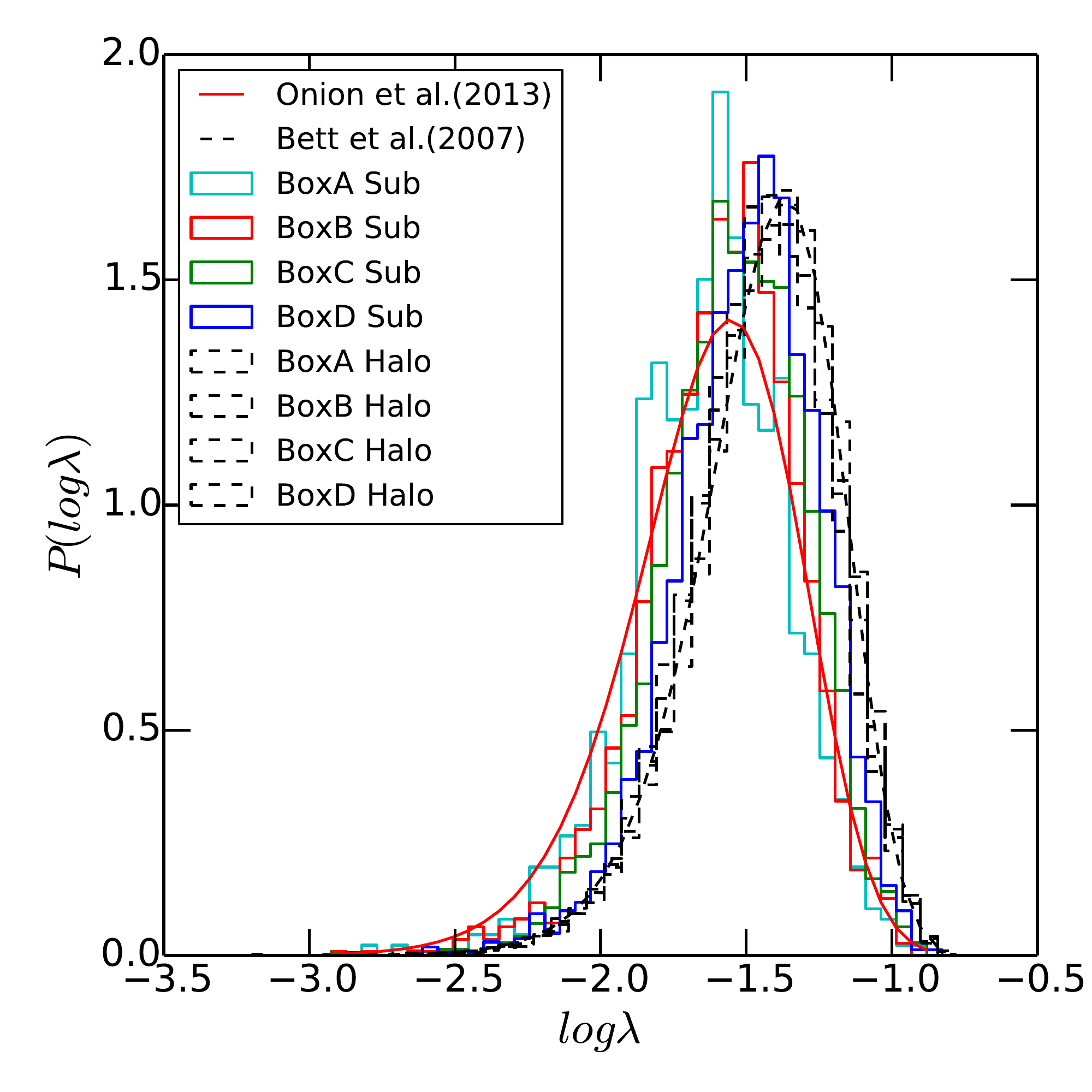}\\
    \includegraphics[width=1\linewidth]{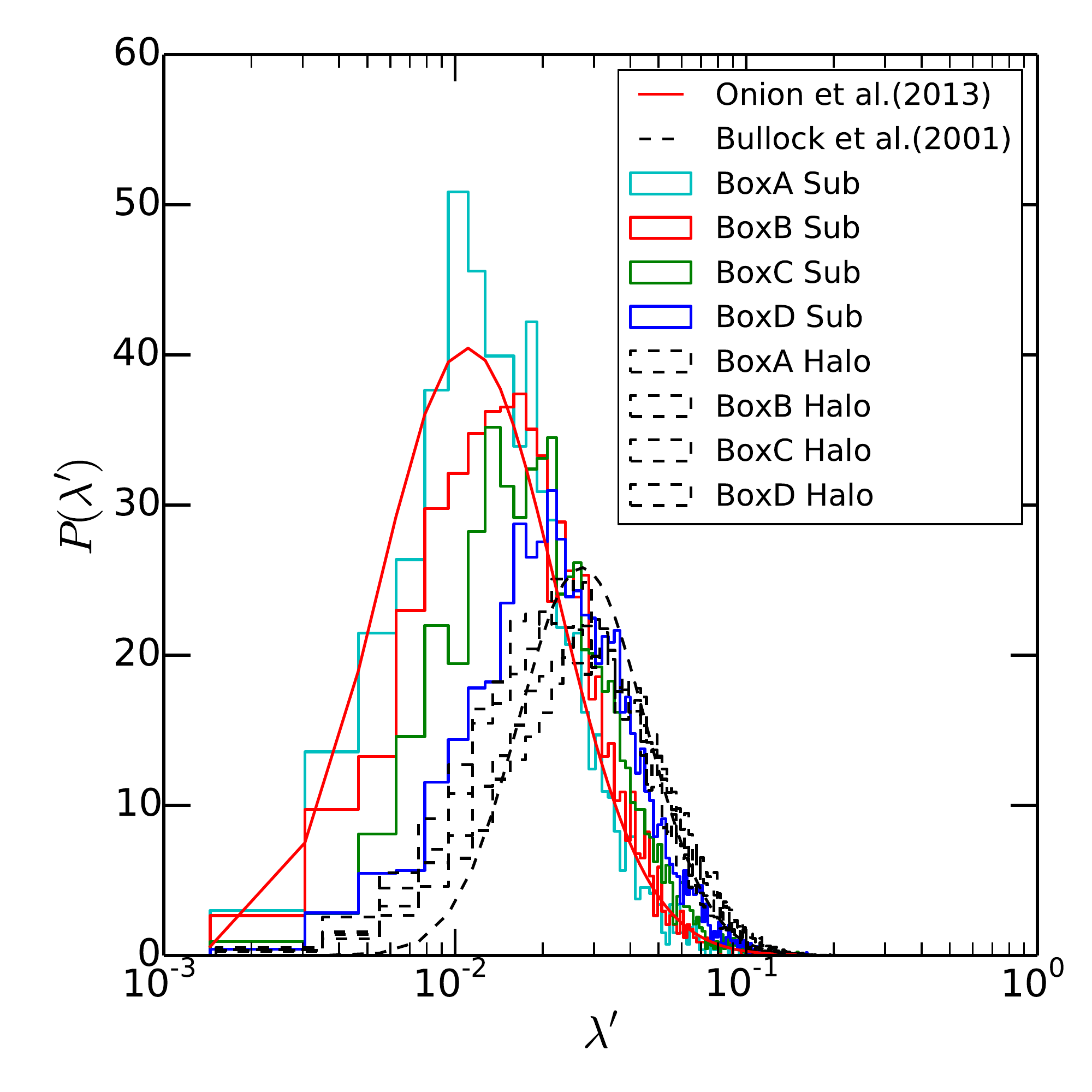}
    \caption{The Peebles (upper) and Bullock (lower) spin distribution of \subfind\ (sub-)\halos\
    at $z=0$. The colored solid histograms show the results for
    sub\halos, and black dashed histogram for \halos. Two lines show
    the best fitting functions from \citet{bullock_2001} (black dashed
    line) and \citet{onions_2013} (red solid line). Only (sub-)\halos\
    with at least 300 particles have been included. For total numbers
    refer to \Tbl{tab:haloes}.}
    \label{pic1}
    \end{figure}

The left and right panel of \Fig{pic1} show the Peebles spin and Bullock spin
distributions of sub\halos\, as well as \halos\, in all the simulation
boxes and the respective fitting functions from \citet{bett_2007},
\citet{bullock_2001} and \citet{onions_2013}. To assure robustness,
only structures resolved with at least 300 particles are included. For
\halos\, we remove particles contained within substructures when we
calculate the spin. We make fits to the histograms in \Fig{pic1}
 using \Eqn{eqn:bettfit} and \Eqn{eqn:bullockfit}. The
respective fitting parameters are given in \Tbl{tab:params}.

    \begin{table*}
     \caption{Parameters for the spin distribution recovered from \subfind\ \halos\ and sub\halos\ in different mass ranges. }
    \centering
    \begin{threeparttable}
    \label{tab:params}
    \begin{tabular}{ l c c c c}
    \toprule
    \multirow{2}{*}{Peebles Spin}&\multicolumn{2}{c}{\halos\ } &\multicolumn{2}{c}{sub\halos\ }\\
    & $\lambda_0$ & $\alpha$ & $\lambda_0$ & $\alpha$ \\
    \hline
    \citet{bett_2007}  & $0.04326\pm0.000020$ & $2.509\pm0.0033$\tnote{*} & - & -\\
    \citet{onions_2013}& - & - & 0.028 & 3.64 \\
    BoxA &$ 0.0371\pm0.00021 $&$ 2.59\pm0.047 $&$ 0.0254\pm0.00049 $&$ 2.92\pm0.17$\\
    BoxB &$ 0.0384\pm0.00020 $&$ 2.63\pm0.043 $&$ 0.0285\pm0.00032 $&$ 2.68\pm0.09$\\
    BoxC &$ 0.0404\pm0.00018 $&$ 2.57\pm0.037 $&$ 0.0308\pm0.00033 $&$ 2.75\pm0.09$\\
    BoxD &$ 0.0411\pm0.00014 $&$ 2.47\pm0.027 $&$ 0.0346\pm0.00030 $&$ 2.54\pm0.07$\\
    \midrule
    \multirow{2}{*}{Bullock Spin}&\multicolumn{2}{c}{\halos\ } &\multicolumn{2}{c}{sub\halos\ }\\
    & $\lambda_0'$ & $\sigma$ & $\lambda_0'$ & $\sigma$ \\
    \hline
    \citet{bullock_2001}& $0.035\pm0.005$& $0.5\pm0.3$\tnote{**} & - & - \\
    \citet{onions_2013} & - & - & 0.018\tnote{***} & 0.70\\
    BoxA &$ 0.0310\pm0.00024 $&$ 0.629\pm0.007 $&$ 0.0167\pm0.00019 $&$ 0.664\pm0.009$\\
    BoxB &$ 0.0332\pm0.00025 $&$ 0.632\pm0.006 $&$ 0.0199\pm0.00020 $&$ 0.669\pm0.008$\\
    BoxC &$ 0.0367\pm0.00028 $&$ 0.622\pm0.006 $&$ 0.0232\pm0.00023 $&$ 0.645\pm0.008$\\
    BoxD &$ 0.0393\pm0.00027 $&$ 0.604\pm0.006 $&$ 0.0283\pm0.00025 $&$ 0.610\pm0.007$\\
    \bottomrule
    \end{tabular}
    \begin{tablenotes}
    \footnotesize
    \item[*] In \citet{bett_2007}, the parameters have much smaller uncertainty because there are much lager populations of halos in the work (17709121 raw FOF halos including 1332239 "clean" ones).
    \item[**] In \citet{bullock_2001}, the parameters have larger uncertainty because they use less haloes for fitting (only 500 haloes).
    \item[***] Note that these parameters differ from the ones originally stated in \citet{onions_2013}. The original values were derived using an incorrect fitting routine. The values stated here are the correct values fitted to the Aquarius L4 data set.
    \end{tablenotes}
    \end{threeparttable}
    \end{table*}

\Fig{pic1}, \Tbl{tab:codes}, \Tbl{tab:res} and
\Tbl{tab:params} all show that the subhalo spin distribution is
different from the halo spin distribution.  This further confirms
earlier results by \citet{onions_2013} who found that the spin
distribution of \emph{sub\halos} in the Level 4 resolution Aquarius
simulation \citep{springel_aquarius_2008} is significantly different
to the one derived by \citet{bett_2007} for \emph{\halos} in the
Millennium Simulation.

Our results reveal new information about the spin of sub\halos.  As
the box size grows from BoxA to BoxD, the discrepancy between
$\lambda_0$ of \halos\ and sub\halos\ decreases gradually, i.e. the
scale of the effect is mass dependent with larger sub\halos\ tending
to have higher spin. One possibility is that in small simulations such
as BoxA or isolated halo models such as Aquarius-A to E studied by
\citet{onions_2013}, large substructures are generally absent.  In the
next section we will demonstrate that subhalo spins increase with
subhalo mass while halo spins do not have a significant mass
dependence.

As an aside, it should be noted that while our fits do not match those
given by \citet{bullock_2001}, \citet{bett_2007} and
\citet{onions_2013} exactly they are within the range of results covered
by these works.  In practice, previous work does not arrive at an
agreement on the exact value of \halos' spin. Most of these studies
fix $\lambda_0'$ in the range $0.031-0.045$, with $\sigma$ between
$0.48-0.64$ \cite[see section 4.1 and Fig.7 in
][]{shaw_2006}. In \citet{bett_2007}, they found median values of
$\lambda_{med}=0.0367-0.0429$ for the entire population of haloes,
depending on the definition of halo,
and $\lambda_{med}=0.043$ for the catalogue of haloes they refined.
Different sets of simulations and halo selection
criteria may result in this variation of the recovered spin
parameter. As \Ssec{Ssec:Test} demonstrated, such factors as mass
resolution and gravitational softening influence the spin.  On the
other hand, the discrepancy between the spin of \halos\ and sub\halos\
within our simulations is much larger than the bias among
simulations. It should therefore be regarded as an intrinsic physical
property rather than a result of different data sets.

\subsection{Mass Dependence}
\begin{figure*}
    \centering
    \includegraphics[width=\linewidth]{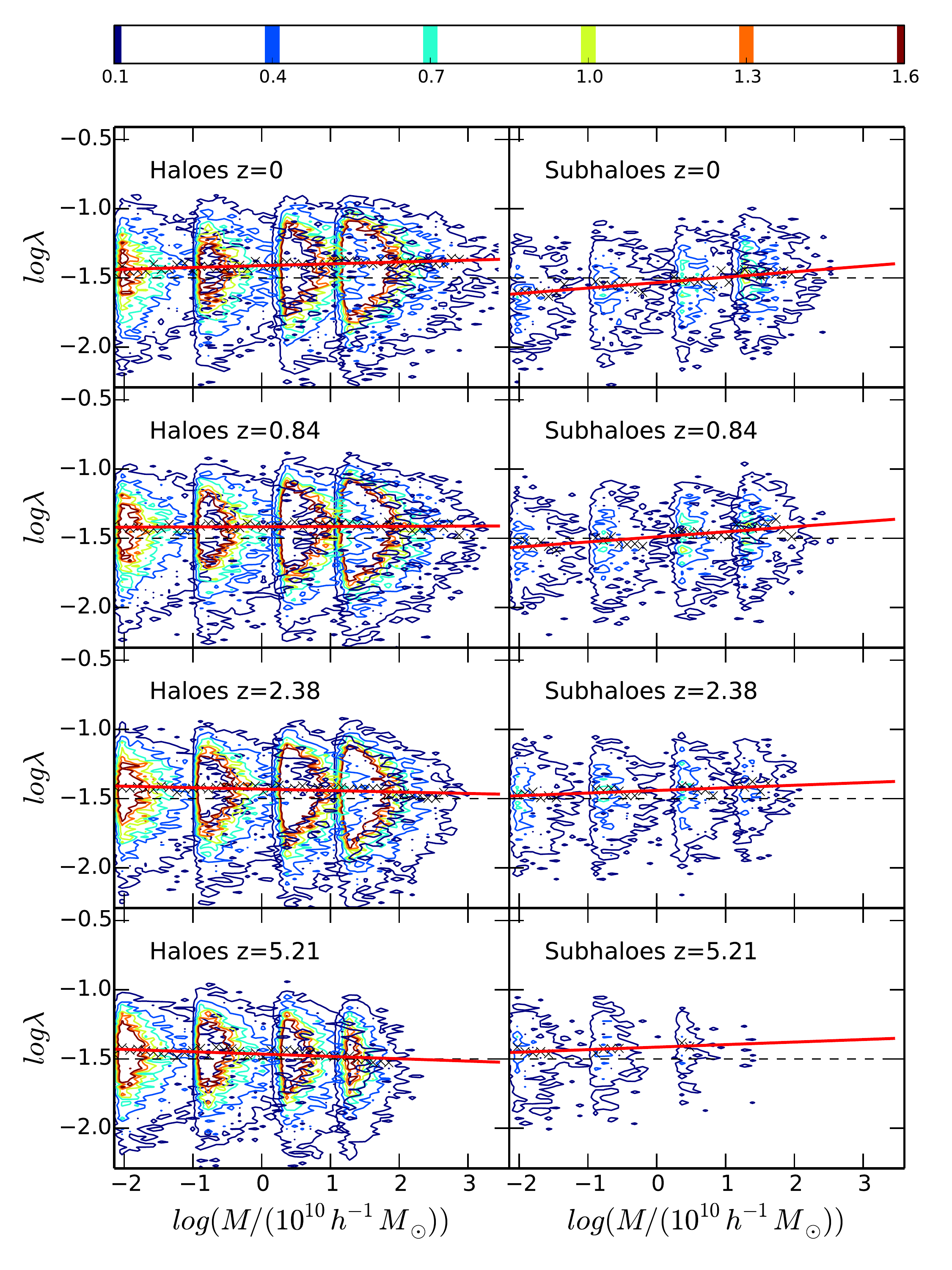}	
    \caption{2 dimensional histogram of Peebles spin against \subfind\ (sub-)\halos\ mass. Contours with different colors represent different level of number density (calculated as $dN/dlog\lambda/dlog(M/10^{10}h^{-1}M_{\odot})/10000$) as indicated in the color bar above. Subplots in left column are statistic for \halos\ and right column for sub\halos. Two plots in a same row are from the same snapshot. The redshift of each row increase from top to bottom respectively. There are in fact four parts of contour in each subplot, which comes from our four simulations respectively. The cross scatters represent $\lambda_0$ in every mass bin. The red thick lines are linear fitting to $\lambda_0$ against mass. The black dashed line indicates a position of $log\lambda=-1.5(\lambda\approx0.032)$ as a standard for comparison. }
    \label{piclm1}
    \end{figure*}

\begin{figure*}
    \centering
    \includegraphics[width=\linewidth]{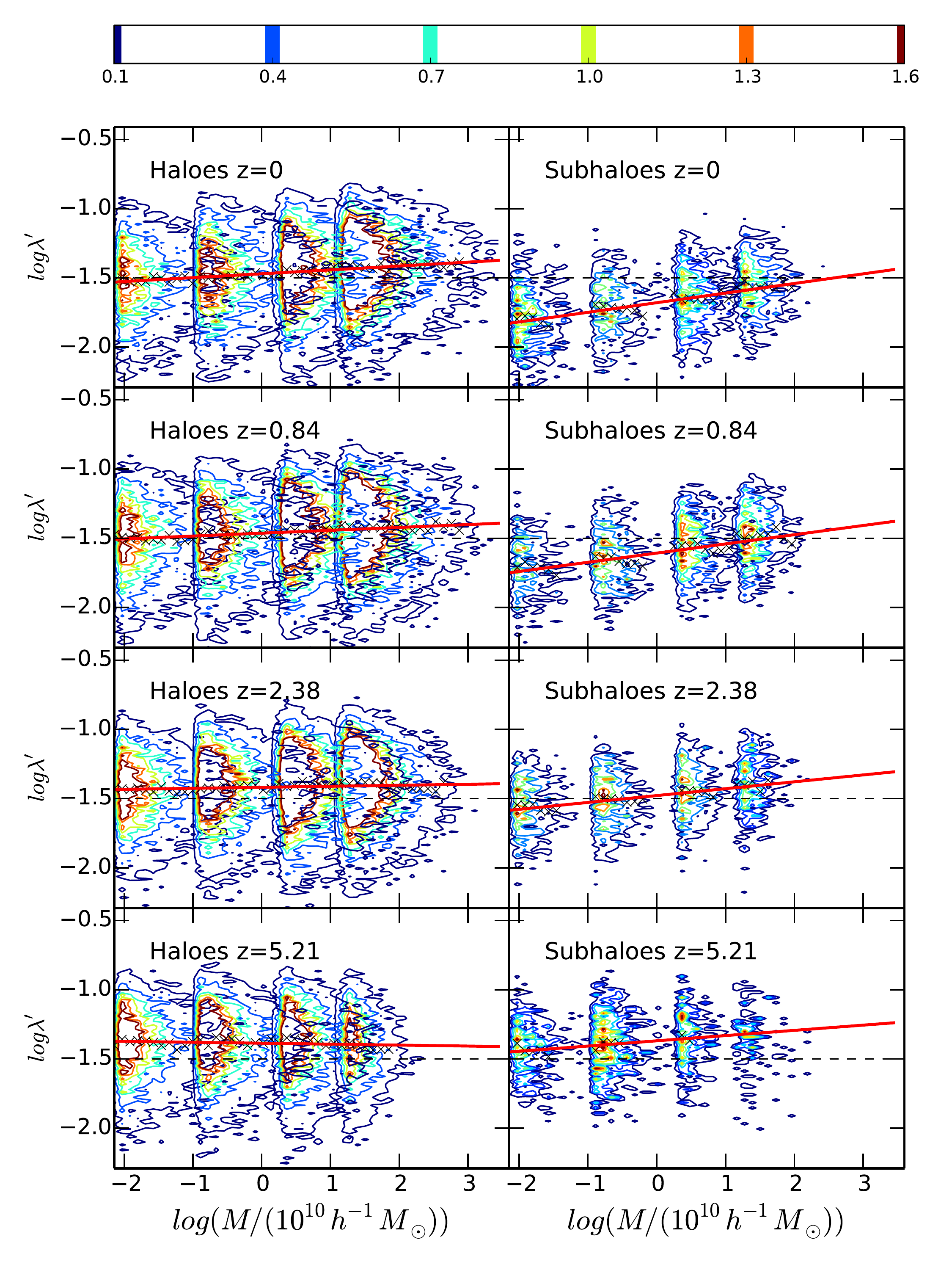}	
    \caption{2 dimensional histogram of Bullock spin against \subfind\ (sub-)\halos\ mass. Contours with different colors represent different level of number density (calculated as $dN/dlog\lambda'/dlog(M/10^{10}h^{-1}M_{\odot})/10000$) as indicated in the color bar above. Subplots in left column are statistic for \halos\ and right column for sub\halos. Two plots in a same row are from the same snapshot. The redshift of each row increase from top to bottom respectively. There are in fact four parts of contour in each subplot, which comes from our four simulations respectively. The cross scatters represent $\lambda_0'$ in every mass bin. The red thick lines are linear fitting to $\lambda_0'$ against mass. The black dashed line indicates a position of $log\lambda'=-1.5(\lambda' \approx0.032)$ as a standard for comparison. }
    \label{piclm2}
    \end{figure*}

As we suggest above, mass dependence can explain the discrepancy
between the spins of \halos\ and sub\halos. To validate this, we
further explore the mass dependence of (sub-)halo
spin. \Fig{piclm1} and \Fig{piclm2} show the two dimensional histogram
of spin against (sub-)halo mass. They present a straightforward
picture of how the spin distribution changes with mass. We use four
simulations to expand the mass range. Contours for each simulation at
the same redshift are normalized and stacked together into one
subplot. We then divide the sample into 40 bins by $log$(sub-)halo
mass, fit the distribution by \Eqn{eqn:bettfit} and
\Eqn{eqn:bullockfit} (if the sample volume in that bin is large
enough). $\lambda_0$($\lambda_0'$) in each mass bin is calculated and
marked on the plots with a cross, and a linear fit to
$\lambda_0$($\lambda_0'$) against mass is indicated by the red solid
line. From the subplots corresponding to redshift 0 (top left and top
right), we can see that, for sub\halos, $\lambda_0$($\lambda_0'$)
clearly increases with increasing subhalo mass. In contrast, the spin
distribution of \halos\ is almost independent of mass. The increasing
subhalo spin with mass is even more pronounced relationship for the
Bullock spin parameter shown in \Fig{piclm2} because for the Bullock
spin the (sub)halo mass has a higher weight
(c.f. \Eqn{eqn:peebles},\Eqn{eqn:bullock}). This results in a larger
discrepancy between \halos\ and sub\halos\ at the low mass end. In
\citet{onions_2013}, their samples are from a Milky Way like
re-simulation, which contains sub\halos\ similar to those found in
BoxA. So the subhalo spin distribution in their work is closer to
that from BoxA. \citet{onions_2013} suggest that the physical
mechanism that drives this difference is mass stripping. Subhalo
particles with high angular momentum are stripped preferentially which
leads to a decrease in the spin parameter. Sub\halos\ with low mass are
usually the ones stripped most severely. Our results strongly support
the claims \citet{onions_2013} made.

The slightly positive slope of $\lambda_{0,halo}(M_{halo})$ and
$\lambda_{0,halo}'(M_{halo})$ is inconsistent with some previous
work. They found that $\lambda_{0,halo}$ is constant or has a slightly
negative slope (\citep[c.f.]{bett_2007, maccio_2007}). However we
should not forget this trend includes the effect of systematic bias
between simulations. Figure 10 in \citet{maccio_2008} shows that
$\lambda_{0}'(M_{halo})$ has a slope of 0.005 in a simulation using
WMAP1, which is the same cosmology as that used here.

\subsection{Radial Dependence}
\begin{figure}
    \includegraphics[width=1\linewidth]{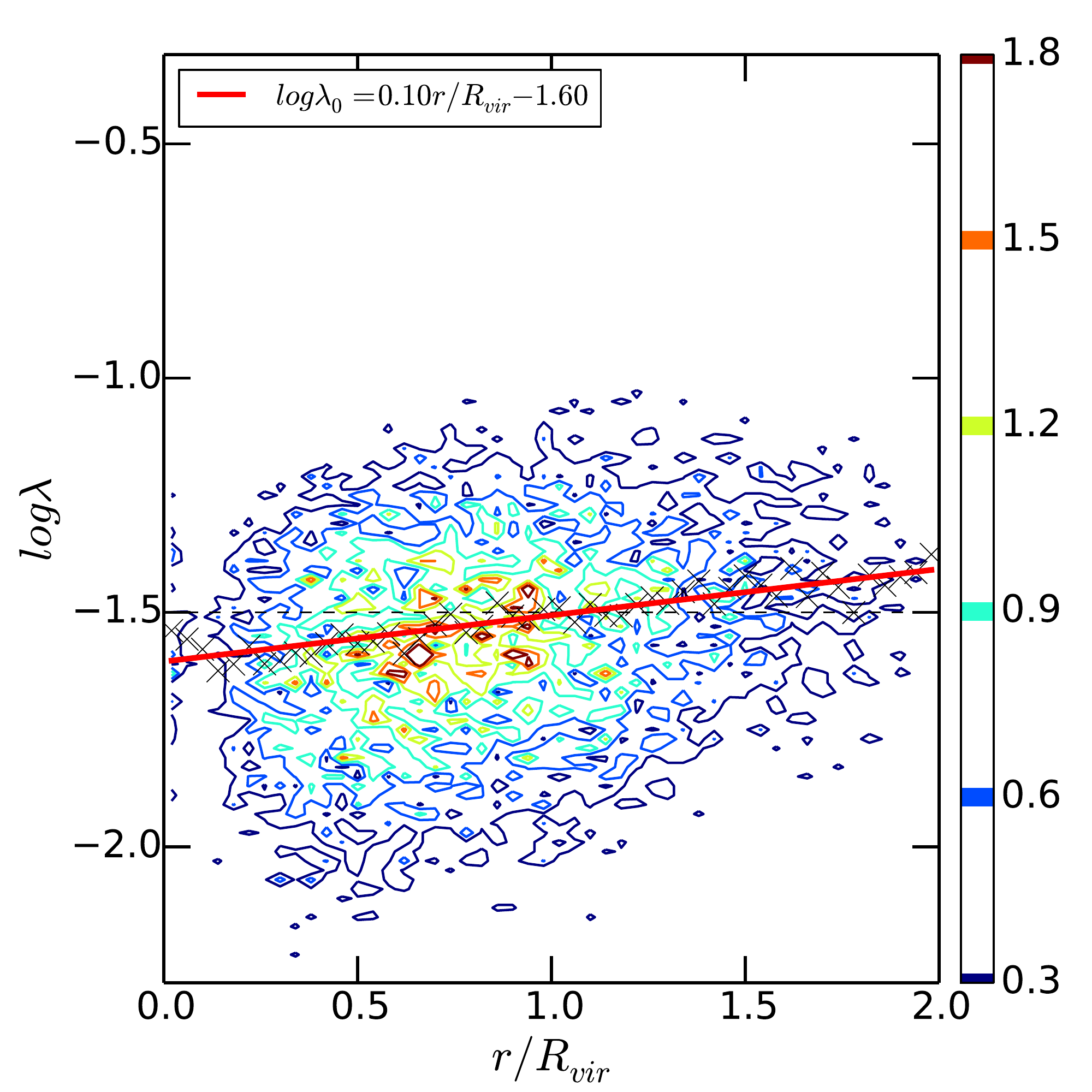}\\
    \includegraphics[width=1\linewidth]{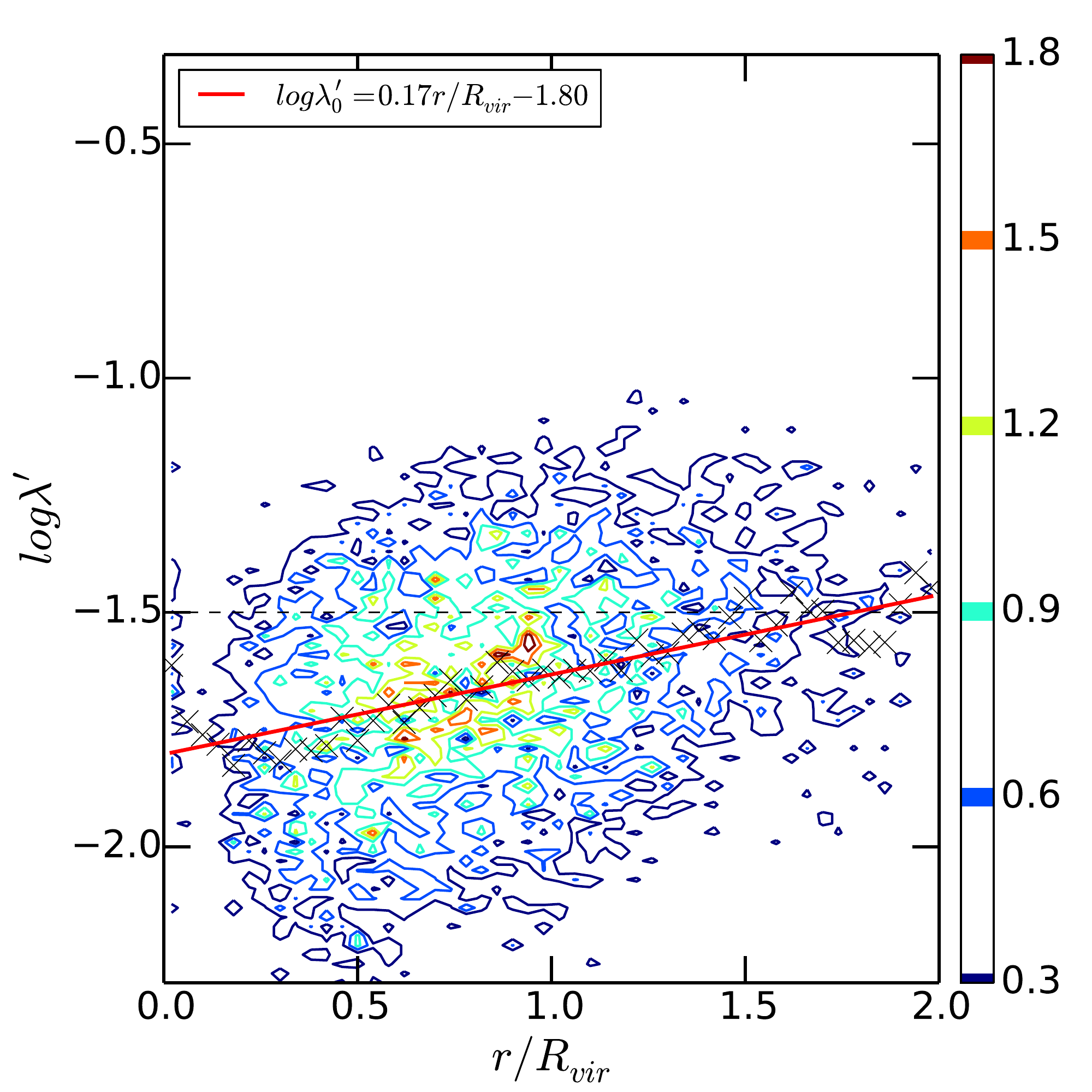}
    \caption{The average Peebles (upper) and Bullock (lower) spin distribution of \subfind\ sub\halos\ against their relative distance from the host halo center. $r$ is the distance of sub\halos\ from the center of their host halo. $R_{vir}$ is the virial radius of host halo. The crosses indicate the peak of every radial bin and the red line is a linear fitting to these marks. Contours with different colors represent different level of number density ( calculated as $dN/dlog\lambda(or \lambda')/dlog(r/R_{vir})/10000$ ) as indicated in the color bars to the right.}
    \label{pic2}
    \end{figure}

To understand further whether tidal stripping of high angular momentum
material could cause the lower subhalo spin distribution, we
investigate the radial dependence of the subhalo spins. Sub\halos\
located closer to the center of their host halo are likely to have
undergone stronger tidal stripping than those nearer the virial
radius. \citet{onions_2013} has done some tests to support their
argument, e.g. they analyze the average spin parameter of sub\halos\
at different distances from the center of the host halo. Here we
perform a more detailed test. We stack subhalo samples from four
simulations together and then make the two dimensional histogram of
spin against their centric distance. Then in each radial bin we fit
the subhalo sample using \Eqn{eqn:bettfit} and
\Eqn{eqn:bullockfit}. Finally we make a linear fit as
$\lambda_0(r)=cr/R_{vir}+d$ or $\lambda_0'(r)=cr/R_{vir}+d$. The
results are displayed in \Fig{pic2} and show that the
spin of sub\halos\ is suppressed close to the center of the host
halo. This is consistent with the argument that sub\halos\ loose their
high angular momentum particles as they are stripped of their outer
layers after infall into a main halo.

\subsection{Redshift Dependence}
So far all our analysis was conducted on the $z=0$ snapshot. However,
the spin distribution is known to change with redshift
\citep{hetznecker_2006}. Hence, we investigated the redshift
dependence of the halo vs. subhalo spin distribution offset. In
\Fig{piclm1} and \Fig{piclm2},we give the Peebles and Bullock spin
distributions of (sub-)\halos\ at $z=0$, $z=0.84$, $z=2.38$ and
$z=5.21$, respectively. The spin of sub\halos\ at the lower mass end
decreases significantly with time, while the spin of massive \halos\
increases slightly with time. We calculate the
$\lambda_0$($\lambda_0'$) in each mass bin and then use a linear
function $log\lambda_0=a*log(M/10^{10}h^{-1}M_{\odot})+b$
($\lambda_0'$ for Bullock spin) to fit $\lambda_0$($\lambda_0'$)
against mass. We list the parameters of each fitting line in
\Tbl{tab:ab}. It is clear the difference between the halo and subhalo
spin distribution increases with time. This is consistent with the
argument that tidal stripping causes the difference. Affected by
stripping, sub\halos\ loose more and more high angular momentum
particles as time passes.

\begin{figure}
    \includegraphics[width=\linewidth]{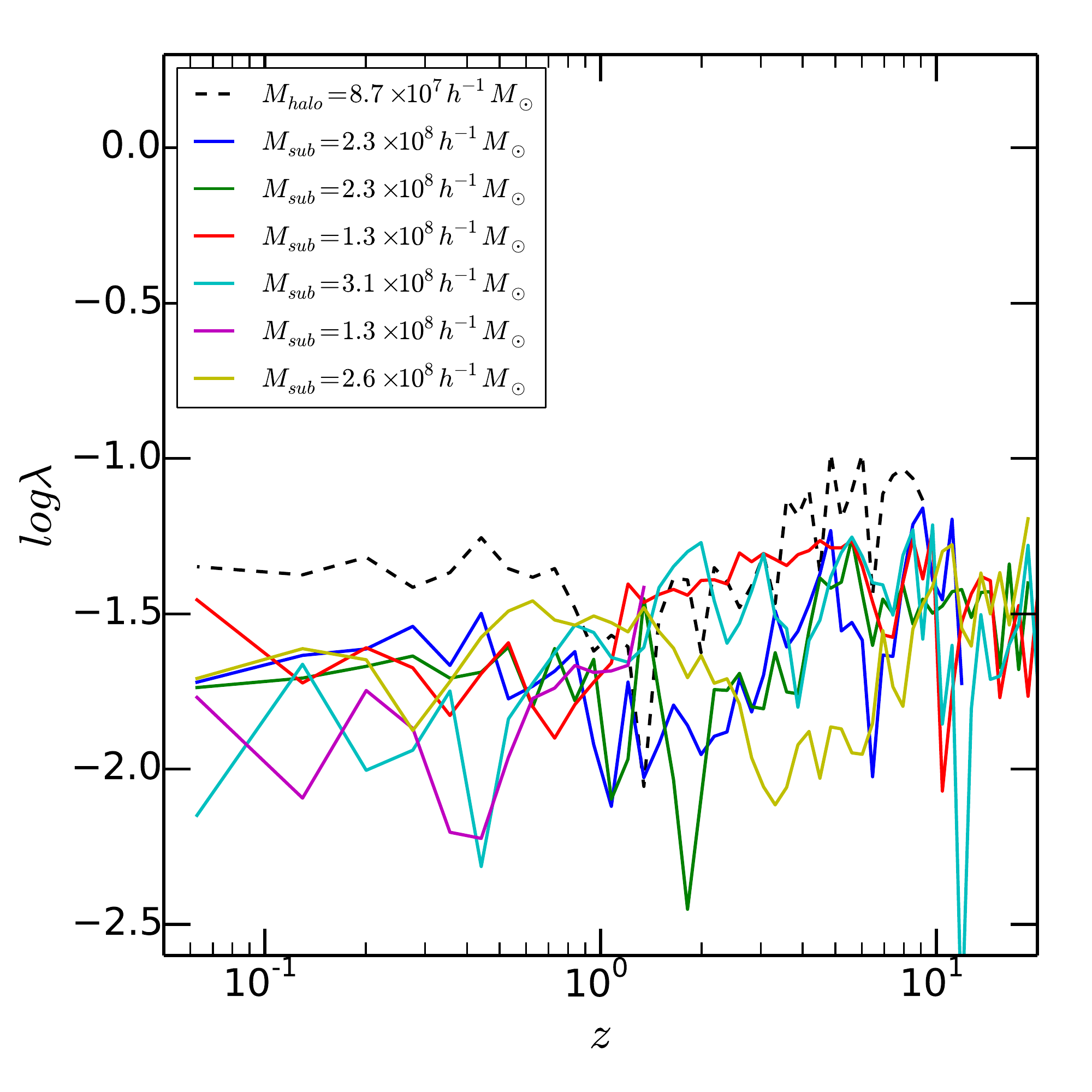}\\
    \includegraphics[width=\linewidth]{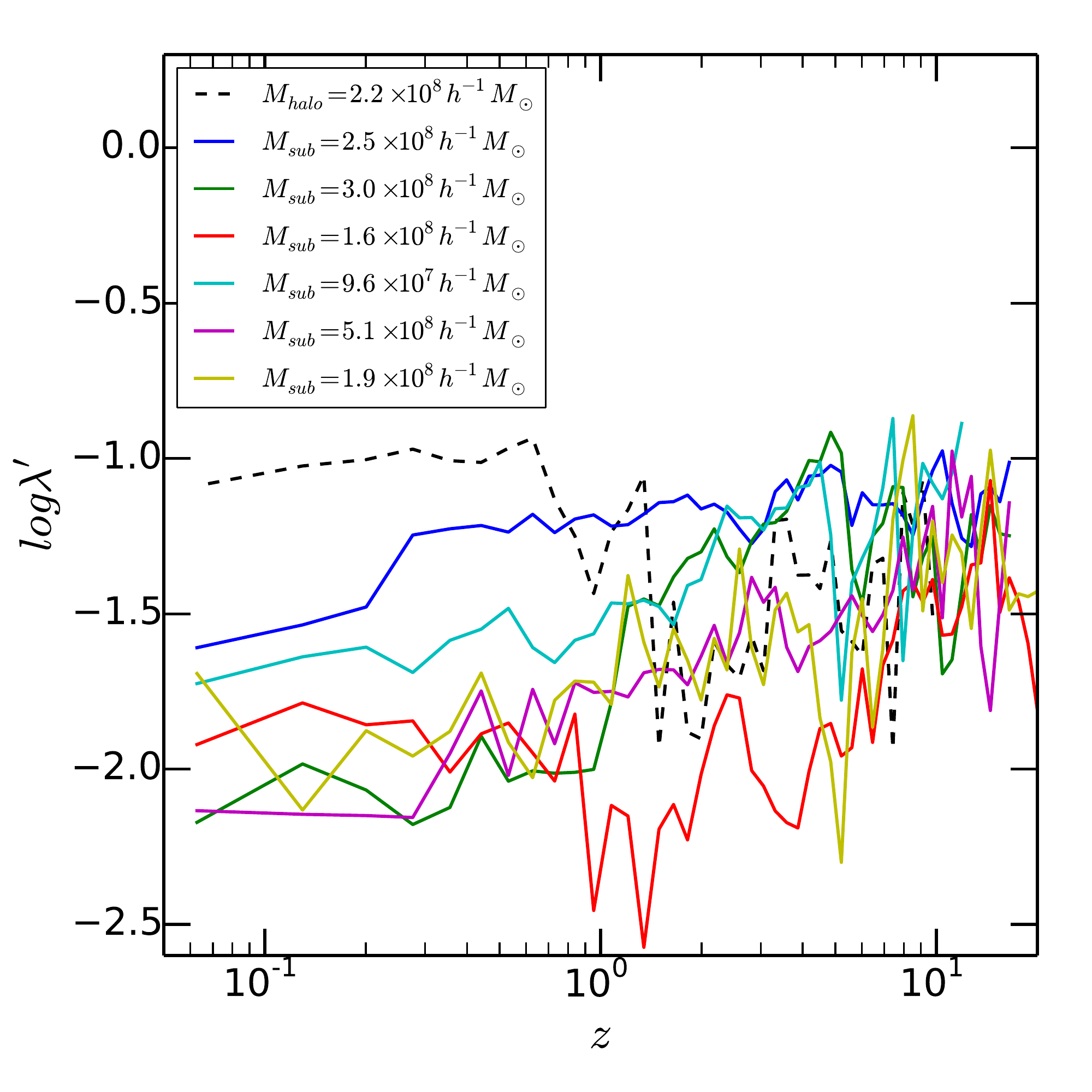}
    \caption{The evolution of Peebles (upper) and Bullock (lower) spin of six selected sub\halos\
    from BoxA. Sub\halos\ have at least 300 particles and are less
    massive than $10^9h^{-1}\rm M_{\odot}$. All selected sub\halos\
    form prior to redshift 8. Solid colored lines represent different
    sub\halos. The black dashed line shows the spin of a halo changing
    with time as a reference.}
    \label{pic3}
    \end{figure}

To confirm this result we checked how the spin of a single halo
changes with time. We randomly select 6 sub\halos\ and plot their spin
against redshift. We constrain the samples selected so that they are likely
to be the sub\halos\ heavily stripped. As expected, heavily stripped
sub\halos\ should have low mass (here we choose masses less than
$10^{9}h^{-1}M_{\odot}$) and have long histories, forming prior to
redshift 8. As shown in \Fig{pic3}, the spin of
sub\halos\ declines at low redshift. We have checked many more
sub\halos\ not displayed in \Fig{pic3} and we find that
most of them display the same trend. We also plot a dashed line for a
halo as reference. The halo is in the same mass range as those
sub\halos\ selected. Its spin almost does not change at low
redshift. This piece of evidence strongly supports the claim that
sub\halos\ suffer from stripping, loosing their spin over time.

The information at the high redshift end in \Fig{pic3}
is not reliable since the progenitors don't contain very many
particles. This results in the large fluctuations seen here.

    \begin{table}
      \caption{Parameters for fitted lines for spin against (sub-)halo mass}
    \label{tab:ab}
    \centering
    \begin{tabular}{ l l  p{1cm}  p{1cm} p{1cm} p{1cm}}
    \toprule
    \multicolumn{2}{c}{ }&\multicolumn{4}{c}{$log\lambda_0=a*log(M/10^{10}h^{-1}M_{\odot})+b$}\\
    \hline
    &&\multicolumn{2}{c}{Halo}&\multicolumn{2}{c}{Subhalo}\\
    &&$a$&$b$&$a$&$b$\\
    \hline
    \multirow{4}{*}{Peebles spin} & $z=0$    & 0.013 & -1.41 & 0.039 & -1.53 \\
                                  & $z=0.84$ & 0.0015 & -1.42 & 0.037 & -1.49 \\
                                  & $z=2.38$ & -0.010 & -1.43 & 0.018& -1.44 \\
                                  & $z=5.21$ & -0.017 & -1.46 & 0.018& -1.41 \\
    \midrule
    \multicolumn{2}{c}{ }&\multicolumn{4}{c}{$log\lambda_0'=a*log(M/10^{10}h^{-1}M_{\odot})+b$}\\
    \hline
    &&\multicolumn{2}{c}{Halo}&\multicolumn{2}{c}{Subhalo}\\
    &&$a$&$b$&$a$&$b$\\
    \hline
    \multirow{4}{*}{Bullock spin} & $z=0$    & 0.028 & -1.47 & 0.069 & -1.68 \\
                                  & $z=0.84$ & 0.021 & -1.46 & 0.066 & -1.61 \\
                                  & $z=2.38$ & 0.0073 & -1.42 & 0.047 & -1.47 \\
                                  & $z=5.21$ & -0.0068 & -1.39 & 0.037 & -1.37 \\
    \bottomrule
    \end{tabular}
    \end{table}



\section{Discussion \& Conclusions} \label{sec:discussion}
In this work we compared the spin distribution function of \halos\ and
sub\halos\ in sets of cosmological box simulations. We found that the
\emph{halo} spin distribution function is well fitted by the
parametrisations given by \citet{bullock_2001} and \citet{bett_2007}
for the Bullock and Peebles \citep{peebles_1969} spin parameter,
respectively.  For the \emph{subhalo} spin distribution function,
however, the typical spin of a small subhalo is significantly lower.
This was previously suggested by \citet{onions_2013} for the
sub\halos\ within Milky Way-like \halos, but is confirmed here for a
full cosmological volume.

We investigated the origin of the difference between the halo and
subhalo spin distributions. We examined the influence of (sub-)halo
finder, spin parametrisation and resolution to confirm that these
factors are not the origin of the difference between the halo and
subhalo spin distributions. In this process we confirmed the
difficulties that the \ahf\ finder has in recovering substructures
reliably where the density contrast between the main halo and the
subhalo is low. We recommend that the \ahf\ halo finder should be
treated with caution in situations where a complete unbiased sample of
the subhalo population is required.

In this paper, we have argued that the difference between the spin
distributions is physical and it is caused by tidal stripping of
sub\halos\ removing high angular momentum material.  This argument is
strongly supported by three pieces of evidence presented here.
Firstly, sub\halos\ tend to have lower spin when compared to \halos\
of the same mass. This discrepancy gets larger towards the low mass
end. Secondly, the spin distribution of sub\halos\ is radially
dependent within a host halo.  Sub\halos\ closer to the host halo
center, which are expected to have been more tidally stripped, have
lower spin than those closer to the virial radius.  Thirdly the
difference between halo and subhalo spin increases with time and hence
is being caused by a dynamical effect, such as tidal stripping.

In summary, we have demonstrated that sub\halos\ typically have lower
spin than \halos\ because tidal stripping removes their highest angular
momentum material.  This can have an important consequence for galaxy
properties that require spin parameter information.  Galaxy properties
that are related to spin are more likely to be correlated to the spin
of the subhalo before infall and not necessarily to its present
value.

\section*{Acknowledgements} \label{sec:Acknowledgements}
This work was supported by the NSFC projects (Grant Nos. 11473053, 11121062, 11233005, U1331201), the National Key Basic Research Program of China (Grant No. 2015CB857001), and the ``Strategic Priority Research Program the Emergence of Cosmological Structures'' of the Chinese Academy of Sciences (Grant No. XDB09010000). YW was supported by the EC framework 7 research exchange programme LACEGAL. HL acknowledges a fellowship from the European Commissions Framework Programme 7, through the Marie Curie Initial Training Network CosmoComp (PITN-GA-2009-238356).  SIM acknowledges the support of the STFC Studentship Enhancement Programme (STEP) and the support of a STFC consolidated grant (ST/K001000/1) to the astrophysics group at the University of Leicester. Part of the simulations in this paper were performed on the High Performance Computing (HPC) facilities at the University of Nottingham (www.nottingham.ac.uk/hpc). This work also made use of the High Performance Computing Resource in the Core Facility for Advanced Research Computing at Shanghai Astronomical Observatory.

The authors contributed to this paper in the following way: YW led the project and is a PhD student of WPL, currently undertaking an extended exchange with FRP.  YW conducted the simulation and analysis with assistance from SIM and JO.  The manuscript was prepared by YW, HL and SIM, with comments and contributions from all authors.

YW also acknowledge Jiaxin Han, Zhaozhou Li for their supports on \hbt.
\bibliography{mn-jour,Haloes} \bibliographystyle{apj}
\label{sec:Bibliography}

circulate
\label{lastpage}

\end{document}